\documentclass[aps,prb,twocolumn,showpacs,amsmath,amssymb,floatfix]{revtex4}
\usepackage{graphicx}
\usepackage{epsfig}
\usepackage{bm}

\begin{document}

\title{Dephasing in the electronic Mach-Zehnder interferometer at filling factor $\nu=2$.}

\author{Ivan P. Levkivskyi$^{1,2}$, Eugene V. Sukhorukov$^1$}
\affiliation{$^1$\
D\'{e}partment de Physique Th\'{e}orique, Universit\'{e} de Gen\`{e}ve,
CH-1211 Gen\`{e}ve 4, Switzerland\\
$^2$\ Physics Department, Kyiv National University, 03022 Kyiv, Ukraine
}
\date{\today}

\begin{abstract}
We propose a simple physical model which describes dephasing in the electronic 
Mach-Zehnder interferometer at filling factor $\nu = 2$. This model explains 
very recent experimental results, such as the unusual lobe-type structure in 
the visibility of Aharonov-Bohm oscillations, phase rigidity, and the asymmetry 
of the visibility as a function of transparencies of quantum point contacts. 
According to our model, dephasing in the interferometer originates from strong 
Coulomb interaction at the edge of two-dimensional electron gas. The long-range 
character of the interaction leads to a separation of the spectrum of edge 
excitations on slow and fast mode. These modes are excited by electron tunneling
and carry away the phase information. The new energy scale associated with the 
slow mode determines the temperature dependence of the visibility and the period 
of its oscillations as a function of voltage bias. Moreover, the variation of 
the lobe structure from one experiment to another is explained by specific charging 
effects, which are different in all experiments. We propose to use a strongly 
asymmetric  Mach-Zehnder interferometer with one arm being much shorter than 
the other for the spectroscopy of quantum Hall edge states.
\end{abstract}


\maketitle

\section{Introduction}
\label{intro}

The quantum Hall effect (QHE),\cite{QHE} one of the central subjects of the 
modern mesoscopic physics,\cite{Datta} continues to attract an attention of 
both experimentalists and theorists. It is well known that the low energy physics 
of the QHE at the Hall plateau is determined by the edge excitations, because at 
strong magnetic fields there exist a gap for excitations in the bulk of the 
two-dimensional gas (2DEG). Properties of quantum Hall edge excitations were 
investigated in a number of experimental and theoretical works.\cite{teor-podborka} 
However, only very recently the progress in the fabrication of novel mesoscopic systems 
made it possible to closely focus on the electronic properties of quantum Hall edge, 
which were not well understood earlier. In particular, experiments on the quantum 
interference and dephasing processes in electronic Mach-Zehnder\cite{Heiblum1}  
interferometers (MZI) brought remarkable results, which shed light on new physics 
of quantum Hall edge states. This physics is the subject of our theoretical investigation. 

\begin{figure}[thb]
\epsfxsize=7cm
\epsfbox{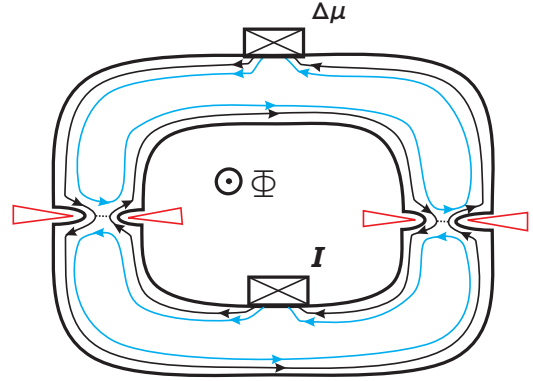} \caption{The Mach-Zehnder interferometer is schematically 
shown as a Corbino disk which contains the two-dimensional electron gas (2DEG). 
In strong magnetic field at filling factor $\nu=2$ two chiral one-dimensional 
channels are 
formed and propagate along the edge of 2DEG. Inner channels (blue line) are 
{\em always} reflected from both quantum point contacts (QPC), while outer 
channels (black line) are mixed by QPCs. Bias $\Delta\mu$ applied to the 
upper ohmic contact causes the current $I$ to flow to the lower ohmic contact. 
This current is due to scattering at QPCs and contains the interference 
contribution sensitive to the magnetic flux $\Phi$ and leading to 
Aharonov-Bohm oscillations.} \vspace{-2mm} 
\label{corbino}
\end{figure}

The idea of the electronic MZI is the same in all recent experiments.
\cite{Heiblum2,Basel,Glattli1,Glattli2,Litvin1} 
The region of the sample, where the two-dimensional electron gas (2DEG) is present, 
is topologically equivalent to so called {\em Corbino disk} (see Fig. \ref{corbino}). 
There are at least two ohmic contacts: one is grounded, and the second is biased by 
the potential difference $\triangle\mu$. The current $I$ is detected at one of the 
ohmic contacts. In fact,   experiments that we discuss used several ohmic contacts 
for the convenience of the measurement, although only two contacts are required for 
the realization of MZI. Two QPCs play a role of beam splitters which mix 
outer edge channels (thin black line in Fig.\ \ref{corbino}). The inner channels 
(blue line in Fig.\ \ref{corbino}) are {\em always} reflected from QPCs.

Typically, the transparencies of two QPCs were varied between $T_\ell=0$ and 
$T_\ell=1$, $\ell = L, R$. However, the most interesting physics was observed 
in two limits: in the regimes of weak tunneling $T_\ell\to 0$ and of weak 
backscattering $T_\ell\to 1$. In the first regime one of the outer channels 
is biased (upper channel in Fig.\ \ref{corbino}) and almost completely  
reflected at the first QPC. Then it runs on the same (upper) part of the 
Corbino disk. The channel that originates from the second (lower) ohmic 
contact is grounded. In the second regime (shown in Fig. \ref{corbino}
as example) 
the biased channels are almost fully transmitted at the first QPC to the 
opposite (lower) part of the Corbino disk. The physical consequences of 
the difference between these two regimes will be discussed later 
in the Sec.\  \ref{correspond}.

Two ohmic contacts are connected solely via scattering at two QPCs. 
Consequently, there are two paths between ohmic contacts, which contribute 
to the total current $I$. The first path is reflected at the right QPC and 
transmitted at the left one, while it is the other way around for the second 
path. It is easy to see that two paths enclose a loop with the nonzero magnetic 
flux. The Aharonov-Bohm (AB) phase associated with it may be changed either by 
varying slightly the strength of the magnetic field, or by varying the length 
of one of the paths with the help of the modulation gate placed near the 
corresponding arm of the interferometer.

According to a frequently used single-particle picture,\cite{Datta} the electron edge
states propagate as plane waves with the group velocity $v_F$ at Fermi level. 
They are transmitted through the MZI (see Fig.\ \ref{corbino}) at the left and right QPCs
with amplitudes $t_L$ and $t_R$, respectively. In the case of
low transmission, two amplitudes add so that the total transmission
probability oscillates as a function of the AB phase $\varphi_{\rm AB}$ and bias $\Delta\mu$. 
The visibility of the oscillations of the differential 
conductance $\mathcal{G} \equiv dI/d\Delta\mu$ is defined as 
\begin{equation}
V_{\mathcal{G}} =\frac{\mathcal{G}_{\rm max}-
\mathcal{G}_{\rm min}}{\mathcal{G}_{\rm max}+\mathcal{G}_{\rm min}}.
\label{vis}
\end{equation} 
Then the Landauer-B\"uttiker formula\cite{Buttiker1} applied to the differential conductance 
gives the following result for the visibility and the AB phase shift:
\begin{equation}
V_\mathcal{G} = \frac{2|t_Lt_R|}{|t_L|^2+|t_R|^2},\hspace{15pt} \Delta\varphi_{\rm AB} 
= \frac{\Delta L}{v_F}\Delta\mu,
\label{nointeraction}
\end{equation}
where $\Delta L$ is the length difference between two paths of the MZI. Thus we arrive at the 
result that in the absence of interaction the visibility is independent of bias, 
while phase shift grows linearly with bias.  

The most remarkable observation made in experiments [\onlinecite{Heiblum2,Basel,Glattli1,Glattli2,Litvin1}]
is that the simple single-particle picture of edge states fails to correctly describe the 
AB effect in the MZI. Essentially, the results can be summarized as following: The visibility of AB 
oscillations is not constant, but rather strongly depends on bias $\Delta\mu$. It oscillates, showing
a new energy scale, and may vanish at specific values of bias. While this behavior is observed in
all experiments, the details are different and very important for understanding the underlying
physics. Therefore, we group experimental observations roughly in two parts, according to a specific
important feature of the experimental set-up, and describe them below in 
details.

\subsection{Only one edge channel is biased}
\label{first}

The first experimental situation that we wish to address is reported in Ref.\ 
[\onlinecite{Heiblum2}]. In this experiment the bias is applied to the outer 
channel only. This situation is achieved by splitting incoming inner and outer 
channels with the help of an additional QPC, so that two channels originate in 
fact from different ohmic contacts. This allows a different bias to be applied 
separately to two channels at the same edge. 

\begin{figure}[h]
\epsfxsize=8cm
\epsfbox{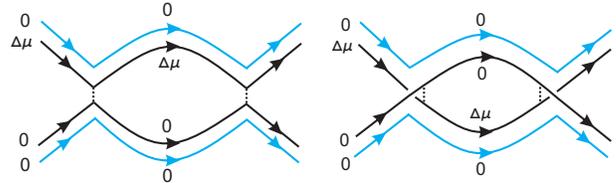} \caption{Schematic representation of the experimental set-up 
in Ref.\ [\onlinecite{Heiblum2}]. Only one edge channel of the MZI is biased. 
Left panel shows the {\em weak tunneling} regime:  Outer edge channels that propagate 
at different arms of the MZI are weakly coupled to each other at two QPCs. Right 
panel shows the {\em weak backscattering} regime: Outer edge channels almost 
completely propagate through QPCs to opposite arms of the MZI and only weakly 
coupled via backscattering. } \vspace{-2mm} 
\label{setup1}
\end{figure}

The MZI in this situation is schematically shown in Fig. \ref{setup1} 
for the regimes of weak tunneling $T_\ell \to 0$ (left panel), and of 
weak backscattering $T_\ell\to 1$ (right panel). This schematics is obtained 
from Fig.\ \ref{corbino} by splitting each ohmic contact attached to the 
Corbino disk and deforming two interfering paths so that they run from left 
to right.  After this procedure, the symmetry between two scattering regimes 
becomes obvious: In order to go from the set-up on the left panel of Fig.\ 
\ref{setup1} to the one on the right panel, one needs to simply flip the 
interferometer vertically. This symmetry is important, and will be  shown 
in Sec.\ \ref{correspond} to result in the symmetry between weak tunneling 
and weak backscattering regimes. 

The Ref. [\onlinecite{Heiblum2}] discovered an unexpected AB effect which 
is inconsistent with the single-particle picture of edge channels. The 
following observations where reported:
\begin{itemize}
\item Lobe-type structure in the dependence of the visibility of AB 
oscillations on the DC bias with almost equal widths of lobes. The 
visibility vanishes at specific values of the bias. This behavior 
persists for various fixed values of magnetic field and for various 
transparencies of QPCs; 
\item The rigidity of the AB phase shift followed by sharp $\pi$-valued 
jumps at the points where the visibility vanishes;
\item The stability of both mentioned effects with respect to changes in the length of 
one of the interferometer paths.
\end{itemize}

The experiment [\onlinecite{Heiblum2}] was theoretically analyzed in several recent 
works [\onlinecite{Sukh-Che,Chalker,Neder,Sim}]. The Ref. [\onlinecite{Sukh-Che}] 
focuses on $\nu=1$ case and suggests that the suppression of the visibility is due 
to the resonant interaction with the counter-propagating edge channel located near 
one of the arms of the interferometer.\cite{footnote1} At present, this idea seems 
to be a reasonable guess, as far as the dephasing at $\nu=1$ is concerned.
However, the experiments [\onlinecite{Heiblum2}] and [\onlinecite{Basel}] 
concentrate on the $\nu=2$ regime, where 
two edge channels coexist. These and new experiments,\cite{Glattli1,Glattli2, Litvin1} where 
the counter-propagating edge channel has been removed, prompt a new theoretical analysis. 
The authors of the Ref.\ [\onlinecite{Chalker}] consider a long-range Coulomb interaction 
at the edge and make an interesting prediction about the temperature dependence of the visibility.
However, they are not able to propose an explanation of the lobe-type behavior of the visibility. 
The Refs.\ [\onlinecite{Neder,Sim}] suggest that dephasing in MZI is due to shot noise
generated by the partition of the edge channel at the first QPC. While this idea may
correctly capture a part of the physics at $\nu=1$, the drawback of this explanation
is that the shot noise vanishes in weak tunneling and weak backscattering regimes,
where the experiments nevertheless demonstrate strong dephasing. Moreover, the experiment
which we discuss below illuminates the special role  
that the second inner edge channel at $\nu=2$ plays in dephasing.

\subsection{Two edge channels are biased}
\label{second}
 
\begin{figure}[h]
\epsfxsize=8.5cm
\epsfbox{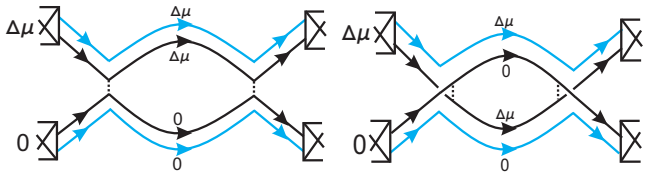} \caption{Schematic representation of the experimental set-up in Ref.\ 
[\onlinecite{Basel}]. Two incoming edge channels of the MZI are biased with the same 
potential difference $\Delta\mu$, and other channels are grounded. Left panel shows 
the {\em weak tunneling} regime, while the right panel shows the {\em weak backscattering} 
regime. } \vspace{-2mm} 
\label{setup2}
\end{figure}

In contrast to the work [\onlinecite{Heiblum2}], the experimental set-up in Ref.\ [\onlinecite{Basel}] 
does not contain an additional QPC that would allow to split two edge channels at $\nu=2$ and to apply 
potentials to each of them separately. Therefore, in Ref.\ [\onlinecite{Basel}] two edge channels that 
originate from the same ohmic contact are biased by the same potential difference $\Delta\mu$. For 
the convenience of the following analysis we again unfold the MZI on Fig.\ \ref{corbino} and 
represent it schematically as shown in Fig.\ \ref{setup2}.

Now it is easy to see the asymmetry between regimes of weak tunneling and of weak 
backscattering. In the first regime (left panel) two channels on the upper arm of 
the interferometer are equally biased with the potential difference $\Delta\mu$. 
The situation is different in the second regime (right panel): The inner channel 
is biased on the upper arm of the interferometer, while the outer channel is 
biased on the lower arm. We believe that this asymmetry is responsible for 
entirely different behavior of the visibility of AB oscillations in the experiment [\onlinecite{Basel}]: 
\begin{itemize}
\item Lobe-type structure with the visibility vanishing at certain values of bias is observed only 
in the weak tunneling regime. The central lobe is approximately two times wider than side lobes. In the
weak backscattering regime the visibility shows oscillations and decays as a function of the bias;
\item No phase rigidity is found at all transparencies of QPCs;
\item The asymmetry in the visibility as a function of the transparency of the first QPC 
is observed. In particular, the visibility always decays as a function of the bias in the 
regimes of weak tunneling. In contrast, in the regime of weak backscattering the visibility 
first grows around zero bias, and only then it decays.
\end{itemize}

It is the last observation which is very important. It indicates that charging effects 
induced by different biasing of edge channels may be  responsible for differences in 
the results of experiments [\onlinecite{Heiblum2}] and [\onlinecite{Basel}]. This 
idea seems to agree with the conclusion of the authors\cite{Roche} of the experiment 
[\onlinecite{Glattli1}]. In this paper we develop this idea and propose a simple model 
that is capable to explain on a single basis all the experimental observations described above.
Namely, we assume a strong (Coulomb) interaction between two edge channels that belong to the same 
quantum Hall edge. The interaction effect is complex: First of all, it leads to charging of 
edge channels and induces experimentally observed phase shifts. Second, the interaction is 
partially screened, which leads to the emergence of the soft mode and of a new low energy 
scale associated with it. The width of lobes in the visibility and the temperature dependence 
are determined by this energy scale. Finally, the interaction is responsible for the decay of 
coherence at large bias.

Further details of our model are given in Sec.\ \ref{model}, while in the Appendix \ref{check} 
we check the consistency of the model. In Sec.\ \ref{visibility} we express the visibility 
of AB oscillations in terms of electronic correlation functions, and derive these 
functions in the Appendix  \ref{correlator}. In section \ref{correspond} we present a 
detailed comparison of our results with the experimental observations. Finally, in 
Sec.\ \ref{conclusion} we briefly summarize our results.

\section{Model of Mach-Zehnder interferometer}
\label{model}

Before we proceed with the mathematical formulation of the model we wish to stress the following points.
The experimentally found new energy scale\cite{Heiblum2,Basel,Glattli1,Glattli2,Litvin1} is very small. For instance,
the width of lobes in the visibility is approximately $20\mu V$. We show below that this energy 
is inverse proportional to the size of the MZI, few micrometers. Thus it is much smaller than any other 
energy scale associated, e.g., with the formation of compressible strips.\cite{strips} 
Therefore, we use an effective model\cite{Wen} appropriate for the  description of the low energy physics
of quantum Hall edge excitations. Namely, we consider the inner and outer edge channels at $\nu=2$
as two chiral boson fields and introduce the Luttinger-type Hamiltonian\cite{teor-podborka,Giamarchi} 
to describe the equilibrium state. Second, we introduce the density-density interaction, 
which is known to be irrelevant in the low-energy limit.\cite{Wen} This fact has no influence on the 
physics that we discuss below, because we focus on the processes at finite energy and length scale, 
which take place inside the MZI.

\subsection{Fields and Hamiltonian}
\label{Hamiltonian}

\begin{figure}[thb]
\epsfxsize=6cm
\epsfbox{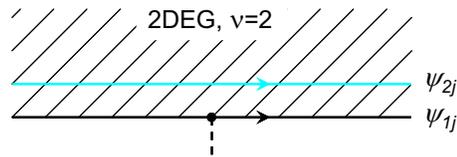} 
\caption{Structure of the quantum Hall edge at $\nu=2$. Two chiral electrons, $\psi_{1j}$ and
$\psi_{2j}$, are propagating along the edge. Tunneling is possible only from and to the 
outer channel ($\psi_{1j}$, black line).} \vspace{-2mm} 
\label{edge}
\end{figure}

We assume that at filling factor $\nu=2$ there are two edge channels at each edge of the quantum Hall system
and two chiral fermions associated with them and denoted as: 
$\psi_{\alpha j}(x)$, $\alpha = 1,2$ and $j=U,D$. Here
the subscript $1$ corresponds to the fermion on outer channel, and 2 to the fermion on inner channel
(see Fig.\ \ref{edge}), while the index $j$ stands for upper and lower arms of the interferometer. 
The total Hamiltonian of the interferometer
\begin{equation}
\mathcal{H}_{\rm tot} = \mathcal{H}_{0} + \mathcal{H}_{\rm int}+\mathcal{H}_T 
\label{hamilt1}
\end{equation}
contains single particle term $\mathcal{H}_{0}$, interaction part $\mathcal{H}_{\rm int}$,
and the tunneling Hamiltonian $\mathcal{H}_T$.

The single-particle Hamiltonian describes free chiral fermions:\cite{Wen} 
\begin{equation}
\mathcal{H}_{0} = 
-iv_F\sum\limits_{\alpha,j} \int dx \,\psi_{\alpha j}^\dagger\partial_x\psi_{\alpha j}, 
\label{hfree}
\end{equation}
where $v_F$ is a Fermi velocity, which is assumed to be the same for each edge channel.
This assumption is not critical, because, as we will see below, the Fermi velocity is 
strongly renormalized by the interaction.

We postpone for a while a detailed discussion of the interaction and at the moment
write the interaction Hamiltonian in terms of local densities $\rho_{\alpha j}$ in the 
following general form:
\begin{equation}
\mathcal{H}_{\rm int} = (1/2)\sum_{\alpha,\beta,j}
\iint dx dy\, U_{\alpha\beta}(x-y) \rho_{\alpha j}(x)\rho_{\beta j}(y).
\label{interaction}
\end{equation}
Note that this effective Hamiltonian is not microscopically derived. However, the experiment
indicates,\cite{Roche} that the interaction has a Coulomb long-range character and leads to
charging effects at the edge. Below we show that once this assumption is made, it leads
to a number of universalities in the MZI physics and correctly captures most of the 
experimental observations.

We have already mentioned in the introduction that the interference in MZI originates 
from scattering processes at QPCs. In the case when interaction is strong, the scattering
has to be assumed weak and treated perturbatively. Fortunately, this limitation does not
detract from our theoretical approach, because neither the interference nor
its suppression are necessarily weak in the case of weak scattering. Moreover, we would like
to stress again that most interesting physics takes place in the regimes of weak tunneling
and of weak backscattering. 

Both regimes can be described by the tunneling Hamiltonian:
\begin{equation} 
\mathcal{H}_{T} = A+A^\dagger\equiv \sum\limits_\ell(A_\ell+A_\ell^\dagger),\quad\ell = L,R,
\label{tunneling}
\end{equation}
where the tunneling amplitude
\begin{equation} 
A_\ell=t_\ell\psi^\dagger_{1D}(x_\ell)\psi_{1U}(x_\ell)
\label{amplitude}
\end{equation}
connects outer edge channels and transfers the electron from the lower arm to the upper
arm of the MZI. It is worth mentioning already here that at low energies the electron
tunneling is relevant and leads in fact to the ohmic behavior of the QPCs, in agreement
with experiments.\cite{Heiblum2,Basel} The AB phase may now be included in the tunneling
amplitudes via the relation $t_R^*t_L = |t_Rt_L|e^{i\varphi_{\rm AB}}$.

\subsection{Bosonization}
\label{bosonization}

In order to account for the strong interaction at the edge, we take advantage of the commonly
used bosonization technique,\cite{Giamarchi} and represent fermion operators in terms 
of chiral boson fields $\phi_{\alpha j}$:
\begin{equation}
\psi_{\alpha j} \propto e^{i\phi_{\alpha j}},
\label{electr}
\end{equation}
which satisfy the commutation relations $[\phi_{\alpha j}(x),\phi_{\alpha j}(y)]=i\pi{\rm sgn}(x-y)$.
The local density is obtained via the point splitting 
\begin{equation}
\rho_{\alpha j}(x) = 
\lim_{\varepsilon\to 0}\psi^\dagger_{\alpha j}(x+\varepsilon)\psi_{\alpha j}(x),
\nonumber
\end{equation}
which gives the following expression:
\begin{equation}
\rho_{\alpha j}(x) = (1/2\pi)\partial_x\phi_{\alpha j}(x).
\label{density}
\end{equation}
Applying point splitting to the single-particle Hamiltonian (\ref{hfree}),
we obtain
\begin{eqnarray}
\mathcal{H}\equiv\mathcal{H}_0 + \mathcal{H}_{\rm int}&=&\sum_{\alpha,\beta, j}
\iint \frac{dx dy}{8\pi^2}\nonumber\\
&\times &V_{\alpha\beta}(x-y)\partial_x\phi_{\alpha j}(x)\partial_y\phi_{\beta j}(y), 
\quad
\label{hamilt2}
\end{eqnarray}
where the interaction potential is simply shifted by the Fermi velocity,
\begin{equation}
V_{\alpha\beta} = U_{\alpha\beta} + 2\pi v_F\delta_{\alpha\beta}\delta(x-y).
\label{V}
\end{equation}
The crucial point is that now the Hamiltonian (\ref{hamilt2}) for quantum Hall edge
is quadratic in boson fields.

Next, we quantize fields by expressing them in terms of 
boson creation and annihilation operators, 
$a^\dagger_{\alpha j}(k)$ and $a_{\alpha j}(k)$,
\begin{eqnarray}
\phi_{\alpha j}(x)& = &\varphi_{\alpha j} + 2\pi p_{\alpha j}\,x\nonumber\\ 
&+& \sum\limits_{k>0} \sqrt{\frac{2\pi}{Wk}}\,[a_{\alpha j}(k)e^{ikx}+a^\dagger_{\alpha j}(k)e^{-ikx}],
\quad
\label{oscill}
\end{eqnarray}
where zero modes, $\varphi_{\alpha j}$ and $p_{\alpha j}$, satisfy commutation relations
$[p_{\alpha j},\varphi_{\alpha j}]=i/W$, and $W$ is the total size of the system. In the end of calculations
we take the thermodynamic limit $W\to\infty$, so that $W$ drops from the final result. 
Then the edge Hamiltonian acquires the following form: 
\begin{eqnarray}
\mathcal{H}&=& (1/2\pi)\sum\limits_{\alpha,\beta,j,k} k\,
V_{\alpha\beta}(k)a^\dagger_{\alpha j}(k)a_{\beta j}(k)\qquad\nonumber\\
&&\qquad\qquad +(W/2)\sum\limits_{\alpha,\beta,j} V_{\alpha\beta}(0)p_{\alpha j}p_{\beta j}
\label{hamilt3}
\end{eqnarray}

The vacuum for collective excitations is defined as $a_{\alpha j}(k)|0\rangle = 0$.
The special care has to be taken about zero modes, because as we show in Sec.\ \ref{correspond}, 
zero modes determine charging effects and phase shifts, which are not small. From the definitions
(\ref{density}) and (\ref{oscill}) it is clear that the zero mode $p_{\alpha j}$ has a meaning
of a homogeneous density at the edge channel $(\alpha, j)$. Therefore, we
define ``vacuum charges'' $Q_{\alpha j}$ 
\begin{equation}
p_{\alpha j} |0\rangle = Q_{\alpha j} |0\rangle, 
\label{vac}
\end{equation}
which are in fact charge densities at the edge channels, generated by the bias.
The energy $E_0$ of the ground state, defined as $\mathcal{H}|0\rangle = E_0|0\rangle$,
is then given by
\begin{equation}
E_0 = (W/2)\sum_{\alpha,\beta, j}
V_{\alpha\beta}(0)Q_{\alpha j} Q_{\beta j}.
\label{electrostat}
\end{equation}

Since edge excitations propagate along the equipotential lines, edge channels can be considered a
metallic surfaces. We therefore can apply the well known electrostatic relation\cite{Landau}
for the potentials $\Delta\mu_{\alpha j}$ to the edge channels:
\begin{equation}
\Delta\mu_{\alpha j}\equiv (1/W)
\delta E_0/\delta Q_{\alpha j}=\sum_{\beta}
V_{\alpha\beta}(0)Q_{\beta j}.
\label{mu}
\end{equation}
Thus the quantity $V_{\alpha\beta}(0)$ is the inverse capacitance matrix.\cite{footnote2}
Using now Eqs.\ (\ref{hamilt3}), (\ref{vac}), and the commutation relation for zero modes,
we arrive at the following important result for the time evolution of zero modes
\begin{equation}
Q_{\alpha j}(t) =\sum_{\beta} V^{-1}_{\alpha\beta}(0)\Delta\mu_{\beta j}, \quad
\varphi_{\alpha j}(t)=-\Delta\mu_{\alpha j}\,t.
\label{Qu}
\end{equation}
 
We finally note that formulated here model of the MZI  is consistent with the effective theory
of the quantum Hall state\cite{Wen} at $\nu=2$. This is demonstrated in the Appendix \ref{check},
where we check the locality of the electron operators, their fermionic commutation relations,
and the gauge invariance of our model.

\subsection{Strong interaction limit and the universality}
\label{univ}

It is quite natural to assume that edge channels interact via the Coulomb potential.
It has a long-range character and the logarithmic dispersion $V_{\alpha\beta}(k)\propto\log(ka)$.
Here $a$ is the shortest important length scale, e.g.\ the width of compressible stripes\cite{strips},
or the inter-channel distance. The dispersion is important in the case $\nu=1$, because it generates
dephasing at the homogeneous edge.\cite{Chalker} However, taken alone the dispersion
is not able to explain lobe-type
behavior of the visibility. What is more important it is the fact that the logarithm may become relatively
large when cutoff at relevant long distances.   

We therefore further assume that the Coulomb interaction is screened at distances $D$, such as 
$L_{U},L_D\gg D \gg a$, where $L_U$ and $L_D$ are the lengths of the arms of the MZI. In fact, some sort 
of screening may exist in MZIs. For instance, in the experiments
[\onlinecite{Heiblum2,Basel,Glattli1,Glattli2,Litvin1}]
the cutoff length $D$ may be a distance to the back gate, or to the massive metallic air bridge. There are
several consequences of screening on the intermediate distances $D$. First of all, it allows to neglect 
the interaction between two arms of the interferometer (see however the discussion in Sec.\ \ref{correspond}).  
Second, at low energies we can neglect the logarithmic dispersion and write
\begin{equation}
V_{\alpha\beta}(x-y) = V_{\alpha\beta}\delta (x-y),
\label{short-range}
\end{equation} 
so that for the Fourier transform we obtain: $V_{\alpha\beta}(k)=V_{\alpha\beta}(0)\equiv V_{\alpha\beta}$. 
And finally, the mutual interaction between inner and outer edge channels, located on the distance of order 
$a\ll D$ from each other, is strongly reduced. 

Therefore, one can parametrize the interaction
matrix  as follows  
\begin{equation}
V_{\alpha\beta}= \pi\left(
\begin{array}{cc}
u+v & u-v \\
u-v & u+v \\
\end{array}
\right),\hspace{12pt}
\label{parametrization}
\end{equation}
where  
\begin{equation}
u/v =\log(D/a)\gg 1,
\label{parameter}
\end{equation}
is new large parameter, the most important consequence of the long-range character
of Coulomb interaction.

Indeed, we now diagonalize the interaction, $V = S^\dagger\Lambda S$, with the result
\begin{equation}
\Lambda = 2\pi\left(
 \begin{array}{cc}
 u & 0 \\
0 & v \\
\end{array}
\right)
,\qquad S = 
\frac{1}{\sqrt{2}}\left(
\begin{array}{cc}
1 & 1 \\
1 & -1 \\
\end{array}
\right).
\label{diag-form}
\end{equation}
Thus we find that the Coulomb interaction at the $\nu=2$ edge leads to the separation 
of spectrum on the fast (charge) mode with the speed $u$ and slow (dipole) mode with
the speed $v$. In Sec.\ \ref{correspond} we show that the lobe structure in the visibility
is determined by slow mode, while the fast mode is not excited at relevant low energies.
That is why at $\nu=2$ the logarithmic dispersion of the Coulomb interaction is not important
for explaining lobes.

 Moreover, the Coulomb character of the interaction leads to the following universality. 
 We show later that the coupling of electrons in the outer channel to the 
 fast and slow mode is determined by the parameters $s_{\alpha} = |S_{1\alpha}| ^2$,
 which satisfy the sum rule 
 \begin{equation}
\sum_\alpha s_{\alpha} = \sum_\alpha |S_{1\alpha}| ^2=1,
\label{unitarity}
\end{equation} 
that follows from the unitarity of the matrix $S$.
For the special choice (\ref{parametrization}) of the interaction matrix 
coupling constants are equal,
\begin{equation}
s_1 = s_2 = 1/2,
\label{limt}
\end{equation}
which has an important consequence, as we show in Sec.\ \ref{visibility}.
Note that in the limit of strong long-range interaction, $u\gg v_F$, 
the result (\ref{limt}) is stable against variations of the bare Fermi velocity 
$v_F$ and is not sensitive to the physics of edge channels at distances 
of order $a$, leading to the universality of dephasing in MZI.

Finally, we partially diagonalize the Hamiltonian by introducing new boson
operators via $a_{\alpha j}(k) = \sum_\beta S_{\alpha\beta}\,b_{\beta j}(k)$. 
Using equations (\ref{hamilt3}), (\ref{parametrization}), and (\ref{diag-form}),
we obtain new Hamiltonian for the quantum Hall edge
\begin{eqnarray}
\mathcal{H}&=& \sum\limits_{j,k}\, [uk\,b_{1 j}^\dagger(k) b_{1 j}(k)+
vk\,b_{2 j}^\dagger(k) b_{2 j}(k)]\qquad\nonumber\\
&&\qquad\qquad +(W/2)\sum\limits_{\alpha,\beta,j} V_{\alpha\beta}p_{\alpha j}p_{\beta j},
\label{hamilt4}
\end{eqnarray}
which completes our discussion of the model. In the Appendix \ref{correlator} we use Eqs.\
(\ref{electr}), (\ref{oscill}), (\ref{Qu}), and (\ref{hamilt4}) 
to derive electronic correlation functions.

\section{Visibility and phase shift}
\label{visibility}

In this section we consider the transport through the MZIs shown in figures \ref{corbino}-\ref{setup2} and 
evaluate the visibility of AB oscillations. Both regimes, of weak tunneling and of weak backscattering, can be 
considered on the same basis, by applying the tunneling Hamiltonian approach.\cite{Mahan}
In the derivation presented below we follow the Ref.\ [\onlinecite{Sukh-Che}]. We introduce 
the tunneling current operator $\hat{I} = \dot N_D=i[\mathcal{H}_T,N_D]$, which differs for two regimes
only by the sign. Here $N_D=\int dx \psi^\dagger_{1D}\psi_{1D}$ is the number of electrons on the 
outer edge channel of the lower arm of the interferometer. Then we use Eqs.\ (\ref{tunneling}) and
(\ref{amplitude}) to write
\begin{equation}
\hat{I} = i(A^\dagger-A).
\label{tunneling-opr}
\end{equation}
We evaluate the average current to lowest order in tunneling and obtain
\begin{equation}
I = \int_{-\infty}^\infty dt \langle[A^\dagger(t),A(0)] 
\rangle \,,
\label{current-av}
\end{equation}
where the average is taken with respect to ground state in quantum Hall edges.
Finite temperature effects will be considered separately in Sec.\ \ref{temperature}.

\begin{figure}[h]
\epsfxsize=4cm
\epsfbox{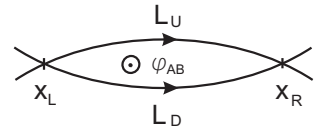} \caption{Schematics of MZI introducing notations: 
$L_U$ and $L_D$ are the lengths of the upper and lower paths of the 
interferometer, respectively. The coordinates of the left and right 
QPC are denoted by $x_L$ and $x_R$, respectively. The magnetic flux 
threading the interferometer results in the AB phase $\varphi_{\rm AB}$.} 
\vspace{-2mm} \label{cnew11}
\end{figure}

It easy to see that the average current can be written as a sum of  four terms: 
\begin{equation}
I=\sum_{\ell,\ell'}I_{\ell\ell'},\quad I_{\ell\ell'}\equiv
\int dt \langle[A_\ell^\dagger(t),A_{\ell'}(0)] \rangle,
\label{current-split1}
\end{equation}
where $I_{LL}$ and $I_{RR}$ are the direct currents at the left and write QPC, respectively,
and $I_{LR}+I_{RL}$ is the interference contribution.
In our model there is no interaction between upper and lower arms
of MZI, therefore the correlation function  in (\ref{current-split1}) splits 
into the product of two single-particle correlators:
\begin{eqnarray}
I_{\ell\ell'} = t^*_\ell t_{\ell'} \int dt 
\hspace{5cm}\nonumber\\
\times  \left[
\langle\psi_{1U}^\dagger(x_\ell,t)\psi_{1U}(x_{\ell'},0)\rangle 
\langle\psi_{1D}(x_\ell,t)\psi_{1D}^\dagger(x_{\ell'},0)\rangle\right.\nonumber \\ 
-\left.\langle\psi_{1U}(x_{\ell'},0)\psi_{1U}^{\dagger}(x_{\ell},t)\rangle 
\langle\psi_{1D}^\dagger(x_{\ell'},0)\psi_{1D}(x_{\ell},t)\rangle\right]
\label{current-split2}
\end{eqnarray}

We note that the operator $\psi_{1j}^\dagger$ applied to the ground state creates a
quasi-particle above the Fermi level (with the positive energy), while the operator
$\psi_{1j}$ creates a hole below Fermi level (with the negative energy). This implies 
that in the first term in (\ref{current-split2}) all the singularities are shifted
to the upper half plane of the complex variable $t$, and in the second term 
singularities are shifted to the lower half plane. This means that only one term contributes,
depending on the sign of bias $\Delta\mu$ which determines the direction of current.
Apart from this, there is no difference between two terms. Therefore, we choose, e.g.,
the first term, shift the counter of integration $C$ to the low half plane, and rewrite
the expression (\ref{current-split2}) as follows:
\begin{eqnarray}
I_{\ell\ell'} = t^*_\ell t_{\ell'} \int_{C} dt \,
\langle\psi_{1U}^\dagger(x_\ell,t)\psi_{1U}(x_{\ell'},0)\rangle \nonumber\\
\times\langle\psi_{1D}^\dagger(x_\ell,t)\psi_{1D}(x_{\ell'},0)\rangle^*,
\label{current-split3}
\end{eqnarray}
where the correlators
are defined in such a way that they have singularities on the real axis of $t$.

The correlators are evaluated in Appendix \ref{correlator} using the bosonization
technique with the result
\begin{eqnarray}
&&i\,\langle\psi_{1j}^\dagger(x_\ell,t)\psi_{1j}(x_{\ell'},0)\rangle 
\nonumber\\
 &&\qquad\qquad\quad =\frac{\exp[i\Delta\mu_{1j}t-2\pi iQ_{1j}(x_{\ell}-x_{\ell'})]}
 {(x_{\ell}-x_{\ell'}-ut)^{s_1}(x_{\ell}-x_{\ell'}-vt)^{s_2}}\qquad
\label{corr-gener}
\end{eqnarray}
One remarkable fact we prove below is that for $x_\ell=x_{\ell'}$ the only
role of the interaction is to renormalize the density of states at Fermi level, 
$n_F=1/(u^{s_1}v^{s_2})$. This immediately follows from the sum rule (\ref{unitarity}).
Therefore, for the direct currents we readily obtain
\begin{equation}
I_{\ell\ell} = 2\pi n_F^2|t_\ell |^2\Delta\mu ,
\label{direct}
\end{equation}
i.e.\ the QPCs are in the ohmic regime, in agreement with experimental observations.

In order to present the visibility in a compact form, we introduce the 
electron correlation functions of an isolated edge, 
normalized to the density of states:
\begin{equation}
G_j(t)=
\frac{\exp[2\pi iQ_{1j}\,L_j]}
 {(t-L_j/u)^{s_1}(t-L_j/v)^{s_2}},\quad j=U,D.
\label{corr-vac}
\end{equation}
This functions contain all the important information about charging effects
(phase shift generated by zero modes), and dephasing determined by the
singularities. Next,
adding all the terms $I=\sum I_{\ell\ell'}$ we find the differential
conductance ${\cal G}=dI/d\Delta\mu$:
\begin{eqnarray}
\mathcal{G} = 2\pi n_F^2(|t_L|^2+|t_R|^2) 
+2n_F^2|t_Lt_R|\,{\rm Im}\Big\{e^{i\varphi_{\rm AB}}\nonumber\\
\times \int_C dt e^{i\Delta\mu t}(t - \Delta t)G_U^*(t)G_D(t)\Big\},
\label{diff-cond}
\end{eqnarray}
where the time shift $\Delta t$ is the charging effect, 
\begin{equation}
\Delta t  = 2\pi\partial_{\Delta\mu}(Q_{1U}L_U - Q_{1D}L_D),
\label{deltat}
\end{equation}
which depends on the bias scheme, and will be calculated in Sec.\ \ref{correspond} 
for a particular experimental situations. It is important to note that in the 
weak backscattering regime (see Figs.\ \ref{setup1} and \ref{setup2}) tunneling occurs
from the lower arm of the interferometer, therefore one should exchange indexes $U$ and $D$.

The first term in Eq.\ (\ref{diff-cond}) is the contribution of direct incoherent currents through
QPCs, while the second term is the interference contribution, which oscillates with magnetic field.
Therefore, the visibility of AB oscillations (\ref{vis}) in the differential conductance $\cal G$
and the AB phase shift take the following form
\begin{equation}
V_\mathcal{G} (\Delta\mu) = V_\mathcal{G} (0)|\mathcal{I}_{\rm AB}|,
\quad \Delta\varphi_{\rm AB} = \arg( \mathcal{I}_{\rm AB}),
\label{visib-def}
\end{equation}
where the visibility at zero bias $V_\mathcal{G}(0)$ is given by Eq.\ (\ref{nointeraction}) 
for a non-interacting system, while all the interaction effects enter via 
the dimensionless Fourier integral 
\begin{equation}
\mathcal{I}_{\rm AB}(\Delta\mu) = \int_C \frac{dt}{2\pi i} \exp(i\Delta\mu t)(t-\Delta t)G^*_U(t)G_D(t),
\label{eta}
\end{equation}
with the counter $C$ shifted to the lower half plane of the variable $t$.
This formula, together with Eqs.\ (\ref{corr-vac}) and (\ref{deltat}) is one of the central
results and will serve as a starting point for the analysis of experiments. However, before we
proceed with detailed explanations of experiments, we would like to quickly consider
two examples. 

\begin{figure}[h]
\epsfxsize=8cm
\epsfbox{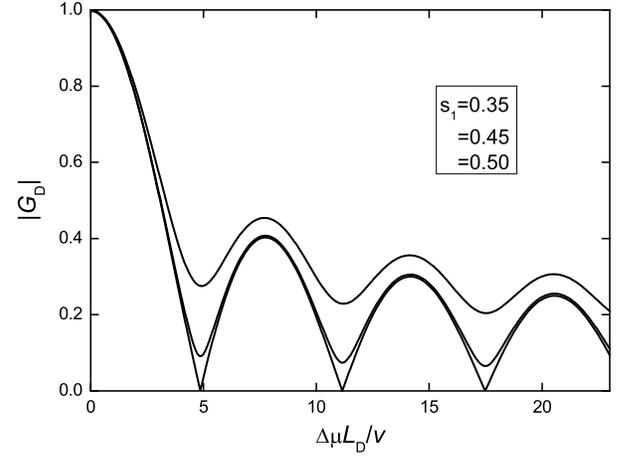} \caption{The absolute value of the Fourier transform 
of the electronic correlation function $G_D(t)$ plotted as a function of the 
dimensionless bias $\Delta\mu L_D/v$ for different values of the 
coupling coefficient $s_1$.} \vspace{-2mm} 
\label{corr}
\end{figure}
 
The first example, a non-interacting system, serves merely as a test for our theory.
In this case using relevant parameters, the vacuum charges $Q_{1U} = \Delta\mu /v_F$,
$Q_{1D} = 0$, the group velocities $u=v=v_F$, coupling constants $s_1=1$, $s_2=0$, we obtain
the correlators $G_U(t) = (t-L_U/v_F)^{-1}\exp(i\Delta\mu L_U/v_F)$ and 
$G_D(t) = (t-L_D/v_F)^{-1}$.  The time shift $\Delta t = L_U/v_F$ follows from Eq.\ (\ref{deltat}). We 
substitute all these results to the Eq.\ (\ref{eta}) and finally obtain:
\begin{equation}
\mathcal{I}_{\rm AB} = \int_C \frac{dt}{2\pi i}  
\frac{e^{i\Delta\mu (t - L_U/v_F)}}{t-L_D/v_F}=e^{i\Delta\mu \Delta L/v_F},
\end{equation}
so that the visibility $|\mathcal{I}_{\rm AB}|=1$, and the phase shift is 
$\Delta\varphi_{\rm AB}=\Delta\mu \Delta L/v_F$, in agreement with the Eq.\ (\ref{nointeraction}).

Next, we consider a more interesting situation when the interferometer is in weak 
tunneling regime (see the Sec.\ \ref{intro}), and one of its arm, e.g.\ the upper arm
of the interferometer, is much shorter than the other, $L_U\ll L_D$. Then the properties
of the  function ${\cal I}_{\rm AB}$ are determined by excitations at the lower arm of MZI at
energies of order $v/L_D$. At this energies the electronic correlator in the upper arm
behaves as a correlator of free fermions: $G_U(t)=1/t$. Therefore, for the visibility
we obtain
\begin{equation}
\mathcal{I}_{\rm AB}= \int_C \frac{dt}{2\pi i}\; e^{i\Delta\mu t}\,G_D(t),
\label{eta2}
\end{equation}
i.e.\ it is simply given by the Fourier transform of the electron correlation function
at the edge. This leads to an interesting idea to use a strongly asymmetric MZI for the 
{\em spectroscopy of excitations} at the edge of quantum Hall system.

We now use the opportunity to analize the role of the coupling coefficients $s_\alpha$
in this simple situation. The absolute value of the Fourier transform of the function 
$G_D$ is shown in Fig.\ (\ref{corr}). We see that $s_1=s_2=1/2$ is the special point. 
In this case, and taking the limit $u\to\infty$, the Fourier transform gives 
$|{\cal I}_{\rm AB}|=|J_0({\Delta\mu L_D/2v})|$,
where $J_0$ is the zero-order Bessel function. Thus the lobes in the visibility of
AB oscillations are well resolved only in the limit of strong long-range interaction.
Therefore, an asymmetric MZI can be used to test the character of the interaction.
From now on we assume that $s_1=s_2=1/2$.

\section{Discussion of experiments}
\label{correspond}

In this section we present a detailed analysis of experiments described in the introduction.
It is convenient to rewrite Eq.\ (\ref{eta}) in slightly different form by 
using Eq.\ (\ref{corr-vac}) with $s_1=s_2=1/2$ and shifting the time integral:
\begin{equation}
\mathcal{I}_{\rm AB}(\Delta\mu) = \oint_C \frac{dt}{2\pi i} 
\frac{t\,\exp(i\Delta\mu t)}{\prod_{j,\alpha}\sqrt{(t+\Delta t-L_j/v_\alpha)}},
\label{eta3}
\end{equation}
where $v_1=u$ and $v_2=v$, and the contour of integration $C$ goes around the branch cuts
(see, e.g., Fig.\ \ref{poles-h}). These branch cuts, which replace single-particle poles of 
correlation functions for free electrons, originate from the interaction. On a mathematical
level, they are the main source of the suppression of the coherence, because at large argument
$\Delta\mu$ the Fourier transform (\ref{eta}) of relatively smooth function quickly decays.
We will use this fact for the analysis of dephasing.  Physically, when electron tunnels, 
it excites two collective modes
associated with two edge channels, and they carry away a part of the phase information. 

On the other hand, charging effects reflected in the parameter $\Delta t$ lead to the bias
dependent shift of the AB phase, $\Delta\phi_{\rm AB}$. As it follows from Eq.\ (\ref{visib-def}),
the phase slips by $\pi$ at points where the visibility vanishes. Away from these points,
in particular at zero bias, the phase shift is a smooth function of the bias. Therefore,
it is interesting to consider the value $\partial_{\Delta\mu}\Delta\phi_{\rm AB}$ at $\Delta\mu=0$
where $|{\cal I}_{\rm AB}|=1$, 
which can be found from the expansion 
$\mathcal{I}_{\rm AB}=|\mathcal{I}_{\rm AB}|e^{i\Delta\phi_{\rm AB}}=
1+i(\partial_{\Delta\mu}\Delta\phi_{\rm AB})\Delta\mu$
in the right hand side of Eq.\ (\ref{eta3}).
We find it exactly:
\begin{equation}
\frac{\partial\Delta\phi_{\rm AB}}{\partial\Delta\mu}=t_0-2\Delta t,\qquad t_0=\frac{u+v}{2uv}(L_U+L_D),
\label{phase-shift}
\end{equation}
where the first term $t_0$ is 
the contribution of the quantum mechanical phase accumulated due to the propagation of an electron
along the MZI. The second term, found from Eq.\ (\ref{deltat}), is the contribution of the charge
accumulated at the arms of MZI due to the Coulomb interaction between edge channels.
Partial cancellation of two effects leads to the phase rigidity found in Ref.\ [\onlinecite{Heiblum2}].
This effect is discussed below. 

Finally, all the experiments found that the visibility $V_{\cal G}$  oscillates as a 
function of the bias $\Delta\mu$. Our model reproduces such oscillations and helps 
to understand their origin. Indeed, two well defined collective modes with speeds
$u$ and $v$ lead to the formation of four branch points in the integral (\ref{eta3}),
which give relatively slowly decaying contributions. These contributions come with different
bias dependent phase factors, so that the function $\mathcal{I}_{\rm AB}(\Delta\mu)$
oscillates. The period of oscillations is determined by the smallest energy scale $\epsilon$, 
which is given by the total size of the branch cut and can be estimated as
\begin{equation}
\epsilon=\frac{2uv}{(u-v)(L_U+L_D)}.
\label{energy-scale}
\end{equation}
In the case $u\gg v$, the parameter $u$ cancels, so that the period of oscillations 
is determined by the slowest mode, and by the size of the interferometer.

We would like to emphasize that oscillations in the visibility appear only when at least 
two modes are relatively well resolved. Our model predicts a power-law decays of the 
visibility. In experiments \cite{Heiblum2,Basel} the visibility seems to decay faster.
There might be several reasons for this, e.g.\ low frequency fluctuations in the
electrical circuit,\cite{Seelig,Buttiker2} or the electromagnetic radiation.\cite{Marquardt1} 
Intrinsic reasons for dephasing deserve a separate consideration. We have already
mentioned that the dispersion of the Coulomb interaction, neglected here, may lead to
strong dephasing.\cite{Chalker}  However, it affects only the fast mode,
while the slow mode contribution to the integral (\ref{eta3}) maintains the phase coherence.
Therefore taken alone the dispersion of Coulomb interaction is not able to explain
strong dephasing at $\nu=2$.  The experiments seem to indicate that the slow mode
is also dispersive, which may be a result of strong disorder at the edge, or,
more interestingly, of the intrinsic structure of each edge channel.\cite{n1}

Having stressed this point, we now wish to focus solely on the phase shift and oscillations
in the visibility. We use the fact that $u\gg v$ and simplify the integral (\ref{eta3})
by neglecting terms containing $1/u$:
\begin{equation}
\mathcal{I}_{\rm AB} = \oint_C \frac{dt}{2\pi i} 
\frac{t\,\exp(i\Delta\mu t)}{(t+\Delta t)\prod_{j}\sqrt{(t+\Delta t-L_j/v)}}.
\label{eta4}
\end{equation}
This expression contains one pole and one branch cut (see Fig.\ \ref{poles-h}).
Therefore, it can be expressed in terms of the zero order Bessel function $J_0$.
After elementary steps we find:
\begin{eqnarray}
\mathcal{I}_{\rm AB} &=& e^{-i\Delta\mu\Delta t}\Big[F(\Delta\mu)-i\Delta t\int_{-\infty}^{\Delta\mu}
d\Delta\mu'F(\Delta\mu')\Big]\nonumber\\
F&\equiv&e^{i\Delta\mu t_0} J_0(\Delta\mu\Delta L/2v),
\label{eta5}
\end{eqnarray}
where $t_0=(L_U+L_D)/2v$, and $\Delta L=L_D-L_U$. 
We now proceed with the analysis of experiments
discussed in the introduction.

\subsection{Only one edge channel is biased}
\label{s-ps}

We start with the experiment [\onlinecite{Heiblum2}]. 
Using Eqs.\ (\ref{Qu}) and
(\ref{parametrization}) we find
\begin{equation}
\left(\!
                          \begin{array}{c}
                            Q_{1j} \\
                            Q_{2j} \\
                          \end{array}
                        \!\right)
= \frac{1}{4\pi uv}\left(
                   \begin{array}{cc}
                     v+u & v-u \\
                     v-u & v+u \\
                   \end{array}
                 \right)\!\left(\!
                          \begin{array}{c}
                            \Delta\mu_{1j} \\
                            \Delta\mu_{2j}\\
                          \end{array}
                        \!\right).
\label{Qus}
\end{equation} 
In the weak tunneling regime, 
shown on the left panel of Fig.\   \ref{setup1}, only outer channel in the upper 
arm of the interferometer is biased, $\Delta\mu_{1U}=\Delta\mu$ and $\Delta\mu_{2U}=
\Delta\mu_{\alpha D}=0$.  
Therefore we obtain
\begin{equation}
Q_{1U}=\frac{u+v}{4\pi uv}\Delta\mu,\quad Q_{1D}=0
\end{equation} 
Then the equation (\ref{deltat}) gives $\Delta t=L_U(u+v)/2uv$.
Substituting $\Delta t$ into Eq.\ (\ref{phase-shift}), we find that
at zero bias
\begin{equation}
\frac{\partial \Delta\phi_{\rm AB}}{\partial\Delta\mu}=\frac{u+v}{2uv}\Delta L.
\label{phase-shift1}
\end{equation}
Therefore, for the symmetric interferometer, $\Delta L=0$, the phase shift
is independent of the bias, away from phase slip points where the visibility
vanishes. This may explain the phenomenon of 
phase rigidity observed in Ref.\ [\onlinecite{Heiblum2}], if we assume
that the interferometer is almost symmetric in this experiment. Indeed, 
the period of oscillations of the visibility is given by the energy scale 
(\ref{energy-scale}). Therefore, the overall
phase shift between zeros of the visibility can be estimated as 
$\Delta L/(L_U+L_D)\ll 1$.

\begin{figure}[h]
\epsfxsize=8cm
\epsfbox{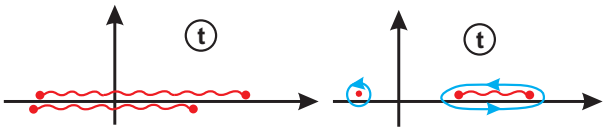} \caption{Analytic structure of the Fourier integral (\ref{eta3}) 
in case of single biased channel.\cite{Heiblum2} {\em Left panel:} Two branch cuts (shown 
apart for convenience) of the integrand come from the product of two single-particle 
correlation functions. {\em Right panel:} In the limit $u \gg v$ two branch points 
corresponding to the fast mode shrink to a single pole at $t=-L_U/2v$, while the slow mode 
produces the branch cut going from $t=L_U/2v$ to $t=L_D/v-L_U/2v$.
The blue line shows the contour of integration $C$.} 
\vspace{-2mm} \label{poles-h}
\end{figure}

\begin{figure}[h]
\epsfxsize=8cm
\epsfbox{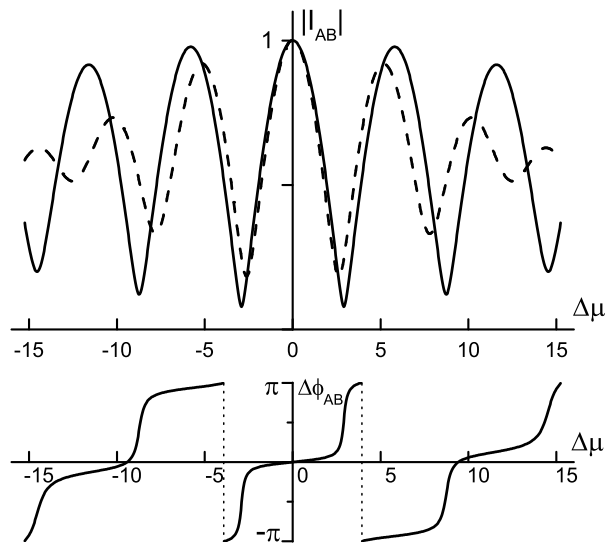} \caption{The intrinsic visibility of AB oscillations 
$|{\cal I}_{\rm AB}|$ and the AB phase shift $\arg( \mathcal{I}_{\rm AB})$ in 
the case of a single biased channel.\cite{Heiblum2} 
{\em Upper panel:} The visibility is plotted as a function of the bias
in units $v/L_U$ for $L_D = 1.15 L_U$ (solid line) and for $L_D = 1.35 L_U$ (dashed line). 
{\it Lower panel:} The phase shift is plotted for $L_D = 1.15 L_U$.}  
\vspace{-2mm} 
\label{visib-h}
\end{figure}

The integral (\ref{eta4}), evaluated numerically, is plotted in Fig.\ \ref{visib-h}
for two values of the asymmetry, $L_D/L_U= 1.15$ and $1.35$. 
Our main focus is first few oscillations of the visibility 
(upper panel), which reveal charging effects. 
We would like to emphasize several points. First, the width of the central lobe 
is equal to the width of side lobes.
This is because in the case of the symmetric
interferometer, $L_U=L_D=L$, the branch cut shrinks to the pole (see Fig.\ \ref{poles-h}),
so that two poles are at $t=\pm L/2v$. Then Eq.\ (\ref{eta4}) gives 
$|\mathcal{I}_{\rm AB}|=|\cos(\Delta\mu L/2v)|$. 
Second, the small variation of the 
length $L_D$ of the lower arm has only minor effect on the position of lobes,
while the amplitude of oscillations is considerably suppressed. Finally,
the lower panel of Fig.\ \ref{visib-h} illustrates the phenomenon of phase 
rigidity for almost symmetric interferometer, $L_D=1.15L_U$. The AB phase shift
changes slowly inside the lobes and slips by $\pi$ at zeros of the visibility.   
All these observation are in agreement with the experiment [\onlinecite{Heiblum2}].

To conclude this section we would like to remark that the visibility in the regime of 
weak backscattering (see the right panel in Fig.\ \ref{setup1}) can be obtained by 
simply replacing $L_U$ and $L_D$. This is because in our model the charging effects 
are important only in the part of the MZI between two QPCs, where they induce phase 
shifts. For the same reason, the transparency of the second QPC does not affect the 
visibility.\cite{Basel} In the next section we show that the symmetry between weak 
tunneling and weak backscattering is broken if the bias is applied to two edge channels.

\subsection{Two edge channels are biased}
\label{f-s}

Next we analyze the experiment [\onlinecite{Basel}]. The details of this experiment 
are discussed in the introduction. In the weak tunneling regime (see the left panel 
of Fig.\ \ref{setup2}) two edge channels
are biased and almost completely reflected at the first QPC. 
Therefore, Eq.\ (\ref{Qus}) gives 
\begin{equation}
Q_{1U} =\frac{\Delta\mu}{2\pi u},\quad Q_{1D}=0.
\end{equation} 
and from Eq.\ (\ref{deltat}) we find $\Delta t=L_U/u$.

\begin{figure}[h]
\epsfxsize=8cm
\epsfbox{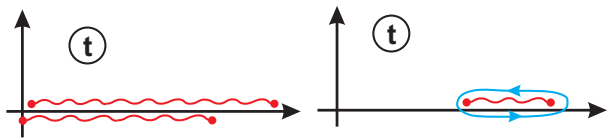} \caption{Analytic structure of the Fourier integral (\ref{eta3}) 
in case when two edge channels are biased,\cite{Heiblum2} and in the weak tunneling regime
(see Fig.\ \ref{setup2}). Left panel shows branch cuts of two single-particle correlation 
functions, while in the right panel the limit $u\gg v$ is taken. The branch cut
extends from $t=L_U/v$ to $t=L_D/v$.} 
\vspace{-2mm} 
\label{poles-s1}
\end{figure}

\begin{figure}[th]
\epsfxsize=8.5cm
\epsfbox{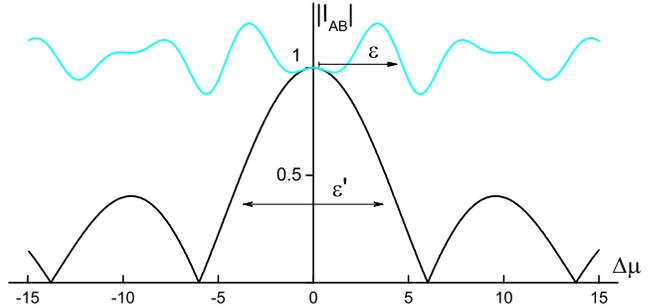} 
\caption{The intrinsic visibility of AB oscillations $|{\cal I}_{\rm AB}|$ 
in the case when two edge channels are biased\cite{Basel} and for strongly asymmetric
interferometer,  $L_D = 1.8L_U$. It is plotted as a function of bias $\Delta\mu$ in units of $v/L_U$
for the regime of weak tunneling (black lime) and for the regime of weak backscattering
(blue line).} 
\vspace{-2mm} 
\label{visib-s2}
\end{figure}

Taking now the strong interaction limit, $u\gg v$, we find that
$\Delta t\to 0$. Therefore, in the integral (\ref{eta4}) the pole 
corresponding to the fast mode cancels (analytical structure
of the integral is shown in Fig.\ \ref{poles-s1}), so that the 
visibility can be found exactly:
\begin{equation}
\mathcal{I}_{\rm AB} =\exp[i\Delta\mu(L_D+L_U)/2v]J_0(\Delta\mu\Delta L/2v),
\label{eta6}
\end{equation}
where $\Delta L=L_D-L_U$.
The visibility of AB oscillations, given by the absolute value of the integral (\ref{eta6}),
is shown in Fig.\ \ref{visib-s2}. One can see that in contrast to the case when only one
channel is biased,\cite{Heiblum2} the central lobe is approximately two times 
wider than side lobes, in agreement with the experimental observation.\cite{Basel}
Moreover, the width of lobes is determined by the new energy scale, $\epsilon'=v/\Delta L$.
Finally, inside the lobes the phase shift $\Delta\phi_{\rm AB}=\Delta\mu(L_D+L_U)/2v$
always grows linearly with bias, so no phase rigidity should be observed.

We now switch to the regime of weak backscattering (see the right panel of Fig.\ \ref{setup2}).
In the upper arm only inner channel is biased, while only outer channel is biased in the lower 
arm of the interferometer. Using again Eq.\ (\ref{Qus}), we obtain
\begin{equation}
Q_{1U} =-\frac{u-v}{4\pi uv}\Delta\mu,\quad Q_{1D} =\frac{u+v}{4\pi uv}\Delta\mu.
\label{Qus2}
\end{equation} 
Then from the equation (\ref{deltat}) we find that $\Delta t = (L_D+L_U)/2v + (L_U-L_D)/2u$. 

\begin{figure}[h]
\epsfxsize=8cm
\epsfbox{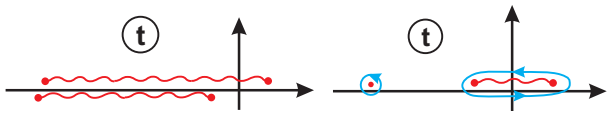} \caption{Analytic structure of the integral (\ref{eta3}), same as
in Fig.\ \ref{poles-s1}, but in the weak backscattering regime. The right panel shows the pole 
at $t = -(L_U+L_D)/2v$ and the branch cut, which extends from $t=-(L_D-L_U)/2v$ to $t=(L_D-L_U)/2v$} 
\vspace{-2mm} 
\label{poles-s2}
\end{figure}

The analytical structure of the integral (\ref{eta4}) is shown in Fig.\ \ref{poles-s2}. 
It looks somewhat similar to the structure shown in Fig.\ \ref{poles-h} for the case of single biased
channel. However, the principal difference between these two cases is that the singularities 
in Fig.\ \ref{poles-s2} are strongly asymmetric with respect to $t\to -t$. In order to see
a consequence of this fact we take the limit $u\gg v$ and write $\Delta t = (L_U+L_D)/2v$. 
For the phase shift (\ref{phase-shift}) at small bias we obtain 
$\partial\Delta\phi_{\rm AB}/\partial\Delta\mu=-(L_U+L_D)/2v$. 
Therefore, in the weak backscattering regime and when two
channels are biased no phase rigidity 
can be observed.  

The most remarkable new feature of the visibility (see Fig.\ \ref{visib-s2})
is that, in contrast to the cases considered above, it grows  as a function of bias 
around $\Delta\mu=0$ , in full agreement with the experiment [\onlinecite{Basel}]. 
It may even exceed
the value $1$ if two QPCs have approximately same transparencies, so that $V_{\cal G}(0)$
is close to $1$. This behavior may look surprising, because it is expected that dephasing
should suppress the visibility of AB oscillations below its maximum value (\ref{nointeraction}) 
for a non-interacting coherent system.  However, one should keep in mind that according to our model 
oscillations of the visibility as a function of bias originate from charging effects which are 
caused by the Coulomb interaction between edge channels. Therefore, simple arguments which
rely on the Landauer formula for the conductance do not apply.

Thus in the experimental set-up where two edge channels are biased\cite{Basel} there is a 
strong asymmetry between weak tunneling and weak backscattering regimes, which is easily seen
in Fig.\ \ref{visib-s2}. In order to clarify the physical origin of this effect, we evaluate 
the integral (\ref{eta4}) in the limit of strong 
interaction $u\gg v$ and for a symmetric MZI, $L_U=L_D=L$. Then the branch cut shrinks to the pole,
and we obtain the following simple result:
\begin{equation}
\mathcal{I}_{\rm AB}=\Delta t/t_0+(1-\Delta t/t_0)e^{i\Delta\mu t_0},
\label{derivative}
\end{equation}
where $t_0=L/v$ is the time of the propagation of the slow mode between two QPCs.
We find that quite similar to the result for the phase shift (\ref{phase-shift}), here we also have a 
competition of two terms, $\Delta t$ given by Eq.\ (\ref{deltat}), and of the flight
time $t_0$. Whether the visibility grows or decays depends on the sign of the second term
in Eq.\ (\ref{derivative}).

In the experiment [\onlinecite{Heiblum2}] $\Delta t=L/2v=t_0/2$, so that the visibility always decays.
On the other hand, the experiment [\onlinecite{Basel}] represent an intermediate case. 
In the regime of weak tunneling we have $\Delta t =0$, while in the regime of weak backscattering
$\Delta t=t_0$, so that in both regimes the visibility is constant for the symmetric MZI.
Therefore, in Fig.\ \ref{visib-s2} we had to consider a strongly asymmetric interferometer
with $L_D=1.8L_U$. Note however, that once $\Delta t$ exceeds slightly $t_0$, the visibility 
easily becomes growing function at small bias. This is exactly what happens if we relax our 
assumption of good screening of the interaction and allow opposite arms of the interferometer 
to interact. Indeed, in order to be electro-neutral the system compensates such interaction by 
decreasing further the charge $Q_{1U}$ below the value given by Eq.\ (\ref{Qus2}), so that now 
$\Delta t>t_0$. We have checked numerically that this assumption alone gives rise to a good 
agreement with the experiment [\onlinecite{Basel}] even in the case
of symmetric interferometer.

\subsection{Effects of finite temperature}
\label{temperature}

The temperature dependence of the visibility of AB oscillations in the MZI has been 
recently measured in Ref.\ [\onlinecite{Glattli2}]. The most interesting fact is that 
the visibility scales exponentially with the total size of the interferometer
$V_{\cal G}\propto e^{-L/l_{\varphi}}$. This is in obvious contradiction
with the prediction $V_{\cal G}\propto e^{-\Delta L/l_{\varphi}}$ for free electrons,
\cite{Buttiker3} where dephasing is due to energy averaging. Moreover, the coherence 
length scales with temperature as $l_\varphi\propto 1/T$, which does not agree
with the prediction based on Luttinger liquid model for $\nu=1$.\cite{Chalker}
Here we show that the experimentally observed temperature dependence of the visibility
can be explained within our model. 

Indeed, according to results of the Sec.\ \ref{visibility}, at high temperatures, 
neglecting charging effects which merely influence the prefactor, 
the visibility can be estimated as $V_{\cal G}\propto\int dt G_D^*(t)G_U(t)$. Here
the correlators are given by the hight-temperature asymptotic form (\ref{step7}),
where $X_\alpha$ has to be replaced with $L_j-v_\alpha t$. Then in the non-interacting 
case (i.e.\ for $s_1=1$, $s_2=0$, and $v_1=v_F$) we obtain the result
\begin{equation}
V_{\cal G}\propto\int dte^{-\pi T\sum_j|t-L_j/v_F|}\propto e^{-\pi T\Delta L/v_F},
\label{T1}
\end{equation} 
which agrees with the prediction in Ref.\ [\onlinecite{Buttiker3}]. On the other hand,
in our model  $s_1=s_2=1/2$, so we obtain
\begin{equation}
V_{\cal G}\propto\int dte^{-\pi T\sum_{\alpha,j}|t-L_j/v_\alpha|}
\propto e^{-(L_U+L_D)/2l_\varphi},
\label{T2}
\end{equation} 
where the dephasing length
\begin{equation}
l_\varphi=\frac{uv}{\pi T(u-v)}.
\label{T3}
\end{equation}
Thus we find that the visibility scales exponentially with the total size of the interferometer,
and the dephasing length scales as $l_{\varphi} \propto 1/T$, in full agreement with the experiment 
[\onlinecite{Glattli2}]. 

Two remarks are in order. According to Eqs.\ (\ref{T2}),  (\ref{T3}), and to the results of the
Sec.\ \ref{visibility}, the temperature dependence and the period of oscillations of the visibility
are determined by the same energy scale $\epsilon$, given by Eq.\ (\ref{energy-scale}).
On the other hand, the decay of the visibility as a function of the bias $\Delta \mu$
at zero temperature is determined by a larger energy scale $\epsilon'$.  It is equal 
to $\epsilon'=v/\Delta L$, or, in case of the symmetric interferometer,  
depends on the dispersion of the slow mode. The existence 
of two distinct energy scales, which originate from the separation of the spectrum
of edge excitations on slow and fast modes, is one of the most important predictions 
of our theory. 

Second, we note that $v$ and $u$ are the group velocities of the collective dipole and charge 
excitations, respectively. Very roughly, they are determined by the spatial separation between edge 
modes $a$, and by the distance to the back gate $D$. On the $\nu=2$ Hall plateau, the 
separation $a$ grows with the magnetic field, because the inner edge channel moves away from the 
edge of 2DEG until it disappears in the end of the plateau.  Therefore, in contrast to the bare 
Fermi velocity, the velocity of the slow mode increases with the magnetic field. 
This may explain the non-monotonic 
behavior of $l_\varphi$ observed in Ref.\ [\onlinecite{Glattli2}]. Indeed,  according
to Eq.\ (\ref{T3}) the decoherence length first increases with the magnetic field starting from
the value $l_\varphi=v/\pi T$. Then it reaches the maximum value at $v\approx u$ and goes down 
to the value $l_\varphi\approx u/\pi T$ on the plateau $\nu=1$.  

\section{Conclusion}
\label{conclusion}

Earlier theoretical works\cite{Buttiker2,Marquardt1,Buttiker3} on dephasing in MZI predicted
a smooth decay of the visibility of AB oscillations as a function of temperature and 
voltage bias. Therefore, when the Ref.\ [\onlinecite{Heiblum2}] reported unusual oscillations
and lobes in the visibility of AB oscillations as a function of bias, this was considered a great
puzzle and attracted  considerable theoretical attention. One of us suggested\cite{Sukh-Che}
a first explanation that is based on the long-range Coulomb interaction between counter-propagating
edge states which leads to resonant scattering of plasmons. Although this phenomenon may be 
encountered in a number of experimental situations, new experiments
\cite{Basel, Glattli1,Glattli2,Litvin1} unambiguously pointed to physics related to the intrinsic
structure of the quantum Hall edge.   

In the present paper we focus on the intrinsic properties of the edge and propose a simple model 
which is able to explain almost every detail of existing experiments. The key ingredient of our
theory is the assumption that two chiral channels at the edge of $\nu=2$ electron system interact 
via the long-range Coulomb potential. This leads to number of universalities, in particular, 
to the separation of the spectrum of edge excitations on slow and fast mode (plasmons), and to equal
coupling of electrons to both modes. When electrons scatter off the QPCs, which play a role
of beam splitters in the electronic MZI, they excite plasmons, depending on the energy provided
by the voltage bias. The plasmons carry away the electronic phase information, which leads
the the decay of the visibility of AB oscillations as a function of bias.    

The remarkable property of our model is that at zero temperature the phase 
information emitted at the first QPC can be partially recollected at the second QPC.
This leads to oscillation and lobes in the visibility which can be interpreted as a
size effect. The new energy scale in these oscillations, associated with the total
size of the MZI and with the slow mode, determines also the temperature dependence 
of the visibility. 

Importantly, within the framework of the same simple model we are able to explain a variety 
of ways the interaction effects manifest themselves in different experiments.
\cite{Heiblum2,Basel, Glattli1,Glattli2,Litvin1}
This includes the lobe-type structure observed in Refs.\ [\onlinecite{Heiblum2,Basel}],
the phase rigidity that was found only in Ref.\ [\onlinecite{Heiblum2}], 
the growing visibility and the asymmetry of the AB effect discovered 
in Ref.\  [\onlinecite{Basel}]. All these phenomena can be interpreted as charging effects.
Indeed, edge channels in quantum Hall systems move along the equipotential lines and
can be regarded as one-dimensional metals. Therefore, they accumulate ground states charges,
which lead to electronic phase shifts, depending on the bias scheme (see figures \ref{setup1} 
and \ref{setup2}). These bias dependent phases determine the overall AB phase shift and 
the specific behavior of the visibility as a function of the voltage bias.  
 
Finally, experimentally observed decay of the visibility as a function of bias seems to be stronger
than what our model predicts. We speculate that this effect cannot be explained by the long-range
Coulomb interaction alone, and may originate from the dispersion of the slow mode due to disorder,
or because of the intrinsic structure of each edge channel.\cite{n1} This point deserves a careful 
experimental and theoretical investigation. Moreover, it is interesting to find out how  
charging and size effects discussed here may influence the interferometry at other filling factors,
where quite similar processes can take place.\cite{Kane} 
Although the first theoretical steps have already been taken,\cite{Gefen,Averin,Ardonne} 
the experiment, as usual, may bring new surprises.

\begin{acknowledgments}

We thank A.\ Boyarsky and V.\ Cheianov for valuable discussions. We are grateful to E.\ Bieri, L.\ Litvin,
S.\ Oberholzer, P.\ Roche, and C.\ Sch\"{o}nenberger for clarifying the experimental details.
This work has been supported by the Swiss National
Science Foundation.

\end{acknowledgments}

\appendix

\section{Consistency of the theory}
\label{check}

Any model of the quantum Hall edge should satisfy the following physical 
conditions:\cite{Wen-Frol} the existence of local electron operator, proper 
charge and statistics of electron operators, and the cancellation of the gauge 
anomaly with the one in the bulk theory. Validity of almost all of them is obvious, 
but it is important to ascertain that there is no intrinsic inconsistencies and
incompatibilities with bulk physics in our theory. In the analysis presented 
below we simplify notations by omitting some indexes, and assuming the summation 
over repeating indexes.

Check of locality of the electron operator (\ref{electr}) is obvious,
\begin{eqnarray}
[\rho_\alpha(x),e^{i\phi_\beta(x')}] =  (1/2\pi)[\partial_x\phi_\alpha(x),e^{i\phi_\beta(x')}]
\nonumber\\ 
= -\delta_{\alpha\beta}\delta(x-x')e^{i\phi_\beta(x')}
\label{check1}
\end{eqnarray}
and follows from the commutation rule for phase operators.
The statistical phase $\theta$ of the operator $\psi_\alpha$ is defined as:
\begin{equation}
\psi_\alpha(x')\psi_\alpha(x) = e^{i\theta}\psi_\alpha(x)\psi_\alpha(x').
\label{check2}
\end{equation}
Using the simple relation 
$$e^{i\phi_\alpha(x')}e^{i\phi_\alpha(x)} = 
e^{-[\phi_\alpha(x'),\phi_\alpha(x)]}e^{i\phi_\alpha(x)}e^{i\phi_\alpha(x')}$$
and the commutation relation for bosonic phase operators
we find that our electron operators (\ref{electr}) are fermions with the phase $\theta = \pi$.
Finally,
the total charge at the quantum Hall edge is 
$$q = \sum\limits_\beta\int dx\,\rho_\beta(x) = 
(1/2\pi)\sum\limits_\beta\int dx\,\partial_x\phi_\beta (x)$$ 
Therefore, using the relation (\ref{check1}) we find
\begin{equation}
[q,\psi_\alpha(x)] = - \psi_\alpha(x),
\label{check3}
\end{equation}
which means that the fermion (\ref{electr}) in our model has an electron charge, $e=1$. 

The only non-trivial question is whether the condition of the cancellation of the anomaly inflow 
imposes any constraint on the interaction matrix $V_{\alpha\beta}$? The answer is no. 
To show this we use the Chern-Simons action for the gauge field $a_\mu$ in the effective
low-energy description of quantum Hall bulk physics \cite{Wen-Frol} at $\nu =2$:
\begin{equation}
S_{\rm CS} = \int dt\int_{\Omega}d^2x\,\varepsilon_{\mu\nu\lambda}a_{\alpha\mu}\partial_\nu a_{\alpha\lambda} \,.
\label{check4}
\end{equation}
Here $\Omega$ is the region of 2DEG where the quantum Hall liquid is present.
After the gauge transformation $a_{\alpha\mu}\to a_{\alpha\mu}+\partial_\mu \lambda_\alpha$ 
the gauge anomaly (total change of action) acquires the following form:
\begin{equation}
\delta S_{\rm CS} = \int dt\int_{\partial \Omega}dx \,\lambda_\alpha(\partial_ta_{\alpha x}-\partial_xa_{\alpha t})
\label{anomaly}
\end{equation}
In our model the action for edge excitations alone can be written as:
$$
S = \int dt\int_{\partial \Omega} dx
(\partial_x\phi_\alpha\partial_t\phi_\alpha - V_{\alpha\beta}\partial_x\phi_\alpha\partial_x\phi_\beta)
$$
The point is that for {\em any} interaction matrix $V_{\alpha\beta}$ the coupling of edge modes with the field 
$a_\mu$ may be written in the gauge invariant form:
\begin{eqnarray}
S(a) &=& \iint_{\partial\Omega}dxdt\,( D_x\phi_\alpha D_t\phi_\alpha \nonumber\\
&-& V_{\alpha\beta}D_x\phi_\alpha D_x\phi_\beta - \epsilon_{\mu\nu}a_{\alpha\mu}\partial_\nu\phi_\alpha)
\label{check5}
\end{eqnarray}
where $D_\mu\phi_\alpha = \partial_\mu\phi_\alpha -a_{\alpha\mu}$. 
After the gauge transformation in the edge action, $\phi_\alpha\to\phi_\alpha +\lambda_\alpha$,
the anomaly (\ref{anomaly}) cancels in the total action $S_{\rm CS}+S(a)$.

\section{Calculation of electron correlation function}
\label{correlator}

After we have introduced the model in Sec.\ \ref{model}, the derivation of the electronic correlation
function is relatively simple. We represent the electronic operators as
$\psi_{1j}\propto \,e^{i\phi_{1j}}$ and fix the normalization in the end of calculations.
Using the gaussian character of the theory, we write 
\begin{equation}
i\langle\psi_{1j}^\dagger(x,t)\psi_{1j}(0,0)\rangle 
\propto \exp(i\Delta\mu_{1j}t-2\pi iQ_{1j}x)K_j(x,t),
\quad
\label{step1}
\end{equation}
where the first term is the average zero mode contribution, while the function $K_j$ 
is the fluctuation part:
\begin{equation}
\log[K_j(x,t)]= \langle[\phi_{1j}(x,t)-\phi_{1j}(0,0)]\phi_{1j}(0,0)\rangle.
\label{step2}
\end{equation} 
Switching to the basis which diagonalizes the Hamiltonian (\ref{hamilt4})
we write:
\begin{eqnarray}
\phi_{1j}(x,t) &=& i\sum_{\alpha,k}\sqrt{\frac{2\pi}{Wk}}\nonumber\\
&\times&\big[
S_{1\alpha}b_{\alpha j}(k)e^{ikX_\alpha} 
+S_{1\alpha}^*b^\dagger_{\alpha j}(k)e^{-ikX_\alpha}\big],\qquad
\label{step3}
\end{eqnarray}
where we introduced the notation $$X_\alpha\equiv x-v_\alpha t$$ 
and neglected fluctuations of zero modes, because we are about to take the thermodynamic 
limit, $W\to\infty$. 
Substituting this expression for the phase operator into the Eq.\ (\ref{step2}),
we obtain,
\begin{eqnarray}
\log[K_j] = \sum_\alpha s_\alpha\int_0^{\Lambda} 
\frac{dk}{k}\left\{n_\alpha(k)(e^{-ik X_\alpha}-1)\right.\nonumber\\
+\left.[1+n_\alpha(k)](e^{ik X_\alpha}-1)\right\} ,
\label{step4}
\end{eqnarray}
where $n_\alpha(k)=[\exp(\beta v_\alpha k)-1]^{-1}$ are the boson occupation numbers,
$\Lambda$ is the ultraviolet cutoff, and $s_\alpha=|S_{1\alpha}|^2$. 

The best way to proceed is to expand occupation numbers in Botlzmann factors,
$n_\alpha(k)=\sum_{m=1}^\infty\exp(-\beta v_\alpha m\,k)$, and integrate each term separately.
This gives 
\begin{equation}
\log[K_j] = -\sum_{\alpha}s_\alpha\!\!\sum_{m=-\infty}^{\infty} 
\log[\Lambda(i\beta v_\alpha m-X_\alpha)].
\label{step5}
\end{equation}
Combining this result with Eq.\ (\ref{step1}), we finally obtain
\begin{eqnarray}
i\langle\psi_{1j}^\dagger(x,t)\psi_{1j}(0,0)\rangle=\exp(i\Delta\mu_{1j}t-2\pi iQ_{1j}x)\nonumber\\
\times
\prod_\alpha\left\{(v_\alpha/\pi T)\sinh[\pi TX_\alpha/v_\alpha]\right\}^{-s_\alpha},
\label{step6}
\end{eqnarray}
where the prefactor
is chosen to be consistent with the free fermion case for $s_1=1$ and $s_2=0$. 
In the zero temperature limit, $T=0$, we obtain the correlator (\ref{corr-gener}).
At high temperatures, $T\gg |X_\alpha|/v_\alpha$ the correlator scales as
\begin{equation}
i\langle\psi_{1j}^\dagger(x,t)\psi_{1j}(0,0)\rangle
\propto \exp\big[-\sum_\alpha \pi Ts_\alpha |X_\alpha|/v_\alpha\big].
\label{step7}
\end{equation}


\begin{thebibliography}{99}

\bibitem{QHE} {\em The Quantum Hall Effect}, edited by R.E. Prange and S.M.
Girvin (Springer, New York, 1987).

\bibitem{Datta}
S. Datta, 
{\em Electronic transport in mesoscopic systems} 
(Cambridge University Press, Cambridge, 1999).

\bibitem{teor-podborka}
For a review, see
A.M. Chang, Rev. Mod. Phys. {\bf 75}, 1449 (2003), and
X.-G. Wen, Adv. in Phys. {\bf 44}, 405 (1995).

\bibitem{Heiblum1} 
Y.\ Ji {\em et al}., Nature (London) {\bf 422}, 415 (2003).


\bibitem{Heiblum2}
I. Neder {\em et al}., Phys. Rev. Lett. {\bf 96}, 016804 (2006).

\bibitem{Basel}
E. Bieri, 
{\em Correlation and Interference Experiments with Edge States}, 
PhD thesis, University of Basel (Oct. 2007);
E. Bieri {\em et al}., to be published.

\bibitem{Glattli1}
P. Roulleau {\em et al.}, Phys. Rev. B {\bf 76}, 161309(R) (2007).

\bibitem{Glattli2}
P. Roulleau {\em et al.}, arXiv:cond-mat/0710.2806.

\bibitem{Litvin1}
L.V. Litvin {\em et al.}, Phys. Rev. B {\bf 75}, 033315 (2007).
L.V. Litvin (unpublished).

\bibitem{Buttiker1}
M. B\"uttiker, Phys. Rev. Lett. {\bf 57}, 1761 (1986).


\bibitem{Sukh-Che}
E.V. Sukhorukov, and V.V. Cheianov, 
Phys. Rev. Lett. {\bf 99}, 156801 (2007).

\bibitem{Chalker}
J.T. Chalker, Y. Gefen, and M.Y. Veillette,
Phys. Rev. B {\bf 76}, 085320 (2007).


\bibitem{Neder}
I. Neder and E. Ginossar, arXiv:0711.1293.

\bibitem{Sim}
S.-C. Youn, H.-W. Lee, and H.-S. Sim, arXiv:cond-mat/0712.2148.

\bibitem{footnote1}
In Fig.\ 1 of the Ref.\ [\onlinecite{Heiblum2}] the counter-propagating edge
state goes from the source S3 to QPC0 and under the air bridge comes
very closely to the upper arm of the interferometer, which is the
part of the channel between QPC1 and QPC2.

\bibitem{Roche}
P. Roche, private communication.

\bibitem{strips}
D.B. Chklovskii, B.I. Shklovskii, and L.I. Glazman,
Phys. Rev. B {\bf 46}, 4026 (1992).

\bibitem{Wen} X.-G. Wen, {\em Quantum Field Theory of Many-Body Systems}
(Oxford University Press, Oxford, 2004).


\bibitem{Giamarchi}
Th. Giamarchi, {\em Quantum Physics in One Dimension}
(Oxford University Press, Oxford, 2003).


\bibitem{footnote2}
Note that $\Delta\mu_{\alpha j}$ is the electrochemical potential, because
our definition (\ref{V}) contains a single-particle contribution.

\bibitem{Mahan}
G.D. Mahan, {\em Many Particle Physics} (Plenum, New York, 1993), 2nd ed.


\bibitem{Landau}
L.D. Landau, E.M. Lifshits, 
{\em Theoretical Physics}, Vol. 8 (Butterworth-Heinemann, Oxford).

\bibitem{Seelig}
G. Seelig, M. B\"uttiker, Phys. Rev. B {\bf 64}, 245313 (2001).

\bibitem{Buttiker2}
G. Seelig {\em et al.}, Phys. Rev. B {\bf 68}, 161310 (2003).

\bibitem{Marquardt1}
F. Marquardt, and C. Bruder, Phys. Rev. Lett. {\bf 92}, 056805 (2004);
Phys. Rev. B {\bf 70}, 125305 (2004).

\bibitem{n1}
J. Fr\"{o}hlich, {\em et al.}, Journ. Stat. Phys. {\bf 103}, 527 (2001).

\bibitem{Wen-Frol}
J. Fr\"{o}hlich, A. Zee, Nucl. Phys. B{\bf 364}, 517 (1991); 
J. Fr\"{o}hlich and T. Kerler, Nucl. Phys. B{\bf 354}, 369 (1991).

\bibitem{Buttiker3}
V. S.-W. Chung, P. Samuelsson, and M. B\"uttiker, Phys.
Rev. B {\bf 72}, 125320 (2005).

\bibitem{Kane}
C.L. Kane, M.P. Fisher, and J. Polchinski, 
Phys. Rev. Lett. {\bf 72}, 4129 (1994).

\bibitem{Gefen}
K.T. Law, D.E. Feldman, and Y. Gefen, Phys. Rev. B {\bf 74}, 045319 (2006). 

\bibitem{Averin}
V.V. Ponomarenko, and D.V. Averin, Phys. Rev. Lett. {\bf 99}, 066803 (2007)

\bibitem{Ardonne}
E. Ardonne, and E.-A. Kim, arXiv:cond-mat/0705.2902; 
B.J. Overbosch, and X.-G. Wen, arXiv:cond-mat/0706.4339; 
B. Rosenow {\em et al.}, arXiv:cond-mat/0707.4474;
W. Bishara, and C. Nayak, arXiv:cond-mat/0708.2704.

\end{thebibliography}
\end{document}